# The Structure Of Barium In The hcp Phase Under High Pressure

W.-S.Zeng*, O.Jepsen

*Max-Planck-Institut für Festkörperforschung, Heisenbergstrasse 1, D-70569 Stuttgart, Germany*

and V.Heine

*Cavendish Laboratory, University of Cambridge, Madingley Road, CB3 0HE, England*

(August 23, 1996)

∗ Present address: Department of Electrical Engineering,

University of Rochester, Rochester, NY 14627, USA.

and Center for Optoelectronics and Imaging,

University of Rochester,

240 East River Rd., Rochester, NY 14623, USA.

# Abstract

Recent experimental results on two hcp phases of barium under high pressure show interesting variation of the lattice parameters. They are here interpreted in terms of electronic structure calculations by using the linear muffin-tin orbital (LMTO) method and generalized pseudopotential theory (GPT) with a nearly free electron plus tight-binding-bond (NFE-TBB) approach. In phase II (5.5-12.6GPa) the dramatic drop in $c/a$ is an instability analogous to that in the group II metals Mg to Hg but with the transfer of $s$ to $d$ electrons playing a crucial role in Ba. Meanwhile in phase $V$ (45-90GPa), the instability decreases a lot due to the core repulsion at very high pressure. The $c/a$ ratio in phase $V$ is somewhat less than the expected ideal value due to some admixture of $5d$ and $4f$ components in the wave function.

# I. INTRODUCTION

The understanding of the physical properties of the heavy alkaline-earth metals has increased significantly with the help of electronic structure calculations and the use of experimental techniques especially at the high-pressure. The $d$-bands in close proximity to the Fermi level and some electronic transfers ( e.g. $s$ to $p$ and $d$, etc. ) are considered as the main reason for the "abnormal" physical properties. At atmospheric pressure barium has a body-centered-cubic ( $bcc$ ) structure and the bottom of the $5d$ band has crossed below the Fermi level [1]. Besides the $sp$-$d$ hybridization, the substantial filling of the $5d$ bands is also very important for the physical properties of barium [2–4] . The filling of the $5d$ band can be increased by the application of high pressure. Under high pressure, an $s$-$d$ electronic transfer may occur due to the crossover of the $6s$ and $5d$ bands. The following high pressure investigations on barium attracted our attention, and the purpose of the present work is to see whether the $s$-$d$ electronic transfer induced by pressure will play an important role and how strong it is.

Takemura [5], by using high pressure ( up to 90GPa ) and powder x-ray-diffraction ( XRD ) techniques, reported that barium at room temperature has *two hcp* ( hexagonal close-packed ) phases under pressure with the following interesting features: *(a)* the $c/a$ ratio of phase $II$ ( in the pressure range: 5.5-12.6GPa) decreases dramatically with pressure, and *(b)* the $c/a$ ratio of phase $V$ ( in the pressure range: 45-90GPa ) is almost pressure-independent with a value of 1.575, which is fairly close to $\sqrt{8/3} \approx 1.632$, the value of ideal $hcp$ structure but yet significant less. As a summary, we show their experimental data in figure 1. A few elements are found to transform back to the same crystal structure under pressure. They are La with $fcc \rightarrow$ distorted $fcc \rightarrow fcc$ [6,7], Cs [8–10] and Ce [11,12] with $fcc \rightarrow fcc$ isostructural transitions. The origins of these transitions are quite different from each other ( see Ref. [7–12] ). The different variation of the $c/a$ ratios as a function of pressure should indicate different stabilizing mechanisms in the two $hcp$ phases of barium.

In this paper, we will interpret the above two interesting features in terms of electronic structure calculations. A linear muffin-tin orbital ( LMTO ) in the atomic-sphere-approximation ( ASA ) [13,14] calculation was performed firstly to obtain the occupation numbers of $s$, $p$, $d$ and $f$ orbitals as functions of pressure in the $hcp$ structure of barium. The electronic transfers can be seen clearly from our LMTO-ASA calculation. The increasing $d$-character in place of $s$ and $p$ favors short bonding length. However, this argument above can not explain the above two features since both $a$ and $c$ lattice constants in the $hcp$ structure become shorter with increasing pressure, as shown in the inset of Fig.1. Thus we have to to use the full-potential LMTO ( FP-LMTO ) [15,16] method to find the relation between the $c/a$ ratio and the total energy of the system. In particular, we will perform some model calculations with different orbitals dropped out of the basis set to investigate their importance. In order to give a good physical explanation of the two interesting features mentioned above, we here used a generalized pseudopotential theory ( GPT ) [17] with a nearly free-electron plus tight-binding-bond ( NFE-TBB ) [18–20] approach in real space to find a relation between instability and the variation of the $c/a$ ratio. The $d$-electron component is found to be the main driving force for the variation of the $c/a$ ratio in phase II. At very high pressure ( e.g. in phase $V$ ), the "hard sphere" nature of the atoms plays the dominant role.

Pseudopotential theory has been used previously to discuss the large variation of the $c/a$ ratio in the series of Be, Mg, Zn, Cd and Hg, both from a real space and reciprocal space points of view [21,22]. The rapid variation of $c/a$ in these series can be described as a developing instability. This is caused by a systematic trend in the pseudopotential associated with decreasing $sp$ hybridization in the series as seen in the variation of $q_0$, the position of first zero pseudopotential in reciprocal space, and an explanation of the variation of the $c/a$ ratio in the $hcp$ structure from Be to Cd can be developed. The $q_0$ in turn can be related to atomic properties such as the radius of the ion core and the $sp$ promotion energy. In barium, because of the substantial filling of $5d$ band as seen from self-consistent LMTO calculations [2–4], the mechanism of the large variation of the $c/a$ ratio is expected to be different. We will show that for barium in phase $II$ the rapid decrease of $c/a$ ratio with pressure is a result of a similar instability but driven by the transfer of $s$ to $d$ electrons.

Other "good" $hcp$ metals have $c/a$ ratio closer to the ideal value. We may expect barium to approach a "hard sphere" behaviour under very high pressure (i.e. in phase $V$) as all materials trend to. Although increased $d$-character will lead to an additional attractive interaction between atoms, the repulsion between atoms at very small interatomic distance corresponding to very high pressure becomes much more important. This will overwhelm the instability apparent in phase $II$ and then create a new stability at larger $c/a$ ratio closer to the ideal value. We will discuss this in detail in section IV.

The paper is organized as follows. Section II will describe brief the calculational methods and some parameters used in this paper. In Section III we will show the results of our LMTO calculation with some discussion. The instability in phase $II$ and the changes of this instability in phase $V$ will be discussed in Sec.IV. The conclusions are given in Sec.V.

## II. DETAILS OF CALCULATIONAL METHODS

In this section, we will briefly describe the calculational methods used in this paper.

We performed self-consistent local density calculation using two version of the LMTO method [13]. The tight-binding-LMTO-ASA ( TB-LMTO-ASA ) version was used to study the changes of the orbital occupations as a function of pressure. This method has been described in details elsewhere [13,23] and a recent detailed application has been published by Jepsen and Andersen [14]. In this method, the one-electron Schrödinger equation is solved selfconsistently in scalar-relativistic scheme. The space is divided into non-overlapping ( small untouching spheres in our FP-LMTO calculation ) or very slightly overlapping ( less than 15% in our LMTO-ASA calculation ) muffin-tin sphere centered at each atom, plus the remaining interstitial region. The potential matrix elements in the spheres are then evaluated by means of expansions in spherical harmonics and their energy derivatives at some set of energies $\varepsilon_\nu$, in the middle of the energy ranges of interest. In the interstitial region, the basis functions are solutions of Helmholtz's equation: $(\nabla^2 - \varepsilon)f(r,\varepsilon) = 0$ with some fixed value of the average kinetic energy $\varepsilon = \kappa_\nu^2$, called as "LMTO envelope". In the LMTO-ASA calculation, the interstitial region is effectively eliminated and only one LMTO envelope with $\kappa_\nu^2 = 0$ is included. We use the orbital-projection technique to get the occupation number of each orbital, projecting the charge onto the $l$ ( $l$=s, p, d and f orbitals ) orbitals and then calculating the projected density of states ( DOS ) $N_l(E)$. The corresponding occupation number $n_l$ of the $l$-orbital can thus be calculated [14]:

$$n_l = \int^{E_F} N_l(E)dE. \tag{1}$$

We used 549 irreducible **k**-points ( from a total of 9216 **k**-points ) to get converged results and a total of 3801 energy mesh points to generate the density of states. The tetrahedron method was used for all **k**-space integrations.

In order to obtain accurate total energy of the system, the potential in the interstitial region was to be considered, properly, not as a muffin tin constant but constructed self-consistently from the wave functions [13]. Thus a triple-$\kappa$ basis set is chosen [15,16] with $\kappa_1^2$=0 $\kappa_2^2$=-1 and $\kappa_3^2$=-2.3 $Ry$ to give a very complete basis set and avoid any geometrical approximation on the potential. In the interstitial region, the basis functions for the wave functions are Hankel functions centered on the atoms. Inside each atomic sphere, each LMTO envelope is a linear combinations of spherical harmonics around the sphere and their energy derivatives at energies corresponding to their principal and orbital atomic quantum numbers. The calculations were within the LDA [24] and scalar-relativistic schemes, using the Hedin-Lundqvist parametrization [25] for the exchange and correlation potential. We included the 5s orbitals as part of the core and treated the 6s,5p,5d,4f as valence states with two sets of energy parameters $\varepsilon_\nu$ appropriate to the pseudocore states and valence states, respectively. By removing some orbitals (especially 5d and 4f) out of the basis set we can investigate their importance. We found that for phase V the whole of the triple-$\kappa$ basis must be included for satisfactory convergence, while for phase II a double-$\kappa$ was found more or less sufficient though the calculated c/a ratio is somewhat larger than the observed value. We found 210 irreducible **k** points ( from a total of 3072 **k**-points ) is sufficient for accurate total energy results. This is consistent with other total energy calculations with the FP-LMTO method for other elements [26–28]. In order to speed up the convergence of the **k**-point sampling, we associated a Gaussian broadening of width 20 mRy with each eigenvalue. We will show the total energy as function of c/a ratio at constant volume corresponding to certain pressures.

For the generalized pseudopotential theory (GPT) calculations, the pair-potential of barium is calculated with the nearly-free-electron-tight-binding-bond (NFE-TBB) method [18–20] to show the instability in c/a. It is not sufficient to use the simple pseudopotential theory of Ref. [21] because of the strong d-component as discussed by Jank and Hafner [29]. The purpose of this part of the work was to give a simple semiquantitative interpretation of the experimental results, not to reproduce them by accurate calculation. Thus the pair potential was not recalculated at each volume since it is slowly varying with electron density [22], and we simply use the pair-potential from Ref. [29] unchanged including both their sp- and d-electron contributions. The distortion compared with the ideal hcp structure is studied by calculating the energy of the system and as a function of the c/a ratio at constant volume, at various volumes corresponding to various pressures in the range covered by the experimental measurements.

## III. LMTO RESULTS FOR PHASE *II* AND *V*

In this section, we will show our LMTO-ASA and FP-LMTO calculational results and some discussions.

Fig.2 shows the calculated occupation numbers of the four orbitals (*s*,*p*,*d* and *f*) as a function of the pressure as calculated with the LMTO-ASA method. The only input parameters in the LMTO-ASA calculations were the lattice constants as shown in the inset of Fig.1. Clearly, the electronic transfer from *s* to *d* in barium distinguishes it compared with the series Mg to Hg [21]. There is clearly a strong transfer from the *s* and *p* orbitals into *d* orbitals with a small transfer into *f* in phase *II*. Additionally in the phase transition from phase *II* to phase *V* there is a larger transfer from *s* to *f*, and then a very small transfer from *d* to *f* with increasing pressure in phase *V*. However, the dominant effect is an *s* to *d* electronic transition induced by pressure all within the same *hcp* structure of barium. The main feature of the density of states for barium under ambient condition is that the bottom of the 5*d* band dips below the Fermi level [1]. Thus the electronic structure contains a substantial filling of the 5*d* bands, as well as the hybridization of the 5*d* states into the 6*s*, 6*p* bands [2–4]. Under pressure the energy of lowest 5*d* band falls further below the Fermi level, even to below the bottom of 6*s* band, due to increasing broadening of the 5*d* band and a drop in its center of gravity. After the same time the 6*s* band rises due to the increase in kinetic energy and being squeezed into the repulsion core, thus contributing to the 6*s* to 5*d* electron transfer. This *s* to *d* transfer under pressure in the alkaline earth metals has been known for a long time [2–4] and is also evident in our calculation ( Fig.2 ). It is a fingerprint of the *fcc* to *bcc* phase transition in several alkaline earth metals [30–32], and indeed can be noted in other elements [2–4] [33]. The 5*d* orbitals are favoured relative to the 6*s*6*p* at smaller interatomic spacings under pressure, and conversely the 5*d* orbitals favour shorter bond lengths. However this argument cannot explain the dramatic drop in the *c/a* ratio in phase *II* since a change of *c/a* from the ideal value increases six bond lengths while reducing six others to the twelve nearest neighbours at constant volume or pressure. Thus the explanation for the drop in $c/a$ in phase *II* ( Fig.1 ) cannot be found in the number of *d*-electrons alone as determined from the LMTO-ASA calculations of Fig.2: it is necessary to go to the FP-LMTO method for full calculation of the total energy as a function of $c/a$. Under very high pressure in phase *V*, the occupied numbers of *s* state is very small ( $\simeq 0.1$/atom ) and the *s* band is effectively emptied as shown in Fig.2. Then the physical properties will be dominated by the repulsion of the *5p* core and incomplete screening of the *z=2* ionic charges as discussed later.

In Fig.3 we show our FP-LMTO calculational results for the total energy as a function of $c/a$ ratio at constant volume chosen to correspond to some of the measured pressure. For these results we used a large basis set to ensure completeness: three *6s* orbitals, three *6p* orbitals, two *5d* orbitals and one *4f* orbital in the triple-$\kappa$ basis sets. The curves in Fig.3 are displaced vertically by arbitrary amounts for clarity. The arrows indicate the experimental $c/a$ ratios, which are seen to agree quite well with the calculated energy minimum. Recently, various other FP-LMTO calculations on the total energy of elements have agreed very well with experimental observations, indicating that the FP-LMTO method is a good and fast calculational method. For example, the structural sequence *bcc*→*hcp*→*fcc*→*bcc* for Cr and Mo and *bcc*→*hcp*→*fcc* for W as a function of pressure were obtained correctly [26], as were the $c/a$ ratios in the *hcp* structure of Ti,Zr and Hf at ambient condition [27]. Similarly, the total energies and bonding features determining the cystallographic structures have been studied in titanium-carbon and tungsten-carbon systems [28], the theoretical investigations comparing fairly well with experiment.

At low pressure in phase *II* ( e.g. 5.91GPa ), the calculated $c/a$ ratio is very close to the observed value. As the pressure is increased, minimum position of the total energy is found to shift rapidly to lower $c/a$ ratio. This corresponds a developing instability in phase *II* and will be discussed in Sec. IV. In phase *V*, the total energy increases steeply with decreasing volume, more than its variation with $c/a$ where equilibrium value remains nearly constant a bit below the ideal value both in experiment and in the calculated results. The calculated energy can be fitted quite well by a simple formula

$$E_{tot} = A(V)((c/a) - (c/a)_{exp})^2 + B(V), \qquad (2)$$

where A(V) and B(V) both increase with decreasing volume. The curves of phase $V$ in Fig.3 are much steeper than those of phase $II$, indicating that the A(V) term in phase $V$ is much larger than that of phase $II$. The physical reason is that, although the $s$-$d$ transfer is in favor of lowering the total energy, the repulsion of the core, the squeezing of the conduction electrons into the core and the incomplete screening of the ionic charge all contribute to a hard-sphere type of behaviour at very small interatomic distance, as will be discussed later.

In order to elucidate the described phenomena further, we will perform a number of model calculations or hypothetical computer experiments which focus on the effects of each orbital. Similar calculations was reported by Ahuja *et al.* on calcium under high pressure [34]. These calculations will provide us a detailed analysis of what features of the electronic structure determine the structural characteristics in $hcp$ barium under pressure. The mechanisms for the drop in $c/a$ ratio in phase $II$ and almost constant $c/a$ ratio in phase $V$ will be found to be different. Our results are shown in Fig.4a ( pressure = 7.21GPa ), typical for phase $II$ and Fig.4b ( pressure = 57.73GPa ), typical for phase $V$. As in Fig.3, each curve is displaced vertically by an arbitrary amount for clarity. The total energy of the system is shown using different basis sets for the valence states, as a function of the $c/a$ ratio at a constant volume corresponding to the experimental volume.

For phase $II$ we first included in the basis set only $s$ and $p$ orbitals as shown in the top curve in Fig.4a. The label *2s2p* means that only $s$ and $p$ orbitals were included in the valence states, two separate $s$ and $p$ orbitals corresponding to two energy eigenvalues $\varepsilon_\nu$ each. The label L=11X refers to the basis set in the interstitial region, the 'one' denoting the highest angular momentum for $\kappa_1$ and $\kappa_2$, and the 'X' denoting that the third set of functions with $\kappa_3$ had been omitted. The second curve labelled L=111, *3s3p* with a larger basis set but still without *d*-orbitals, is almost parallel to the first so that we may consider our results converged, within the stated restriction of course. Note that the minimum is at *c/a* larger than ideal ( 1.632 ) as it is in Zn and Cd with predominantly *sp*- electrons. The other curves in Fig.4a shows that adding one *d*-orbitals inside the muffin tin sphere reduces *c/a* to 1.59, i.e. less than for the ideal structure, and two much *d*-orbitals reduces it further to the experimental value of 1.54 at 7.21 GPa. The inclusion of *f*-orbitals inside and outside the spheres does not affect the results significantly.

Similar calculation for phase *V* are shown in Fig.4b. Again the exclusion of *d*-orbitals in the basis set results in *c/a* larger than ideal, and the inclusion of two *d*-orbitals brings it down to the measured value 1.57, less than the ideal. However in each case the *c/a* is closer to the ideal value than in Fig.4a, indicated a " hard sphere " type of situation as might be expected at much high pressure.

## IV. BEHAVIOUR OF c/a IN PHASES *II* AND *V*

In this section, we will consider the instability of the *c/a* ratio in phase *II* and the short range repulsion in phase *V*.

This instability can best be understood in terms of pseudopotentials and second order perturbation theory. This was used to explain the instability in the series Be to Hg [35] and we will take it over to apply to barium with one modification. In this theory, the total energy of the system for arbitrary position $R_{ij}$ of the atoms can be written [35,19]

$$E_{tot} = f(V) + \sum_{ij} \Phi(R_{ij}, V) \qquad (3)$$

where the dominant term f(V) depends on the volume only and includes the main part of the kinetic, exchange and correlation energy of the electron gas and of the electrostatic energy of the ion core in the electron gas. For a given pressure, it largely controls the resulting volume. The $\Phi(R_{ij}, V)$ is a spherically symmetric pair potential between the ions, not strongly dependent on the volume V of the system [22] [36], so that for brevity we shall not indicate the dependence on V explicitly hereafter. Note that $\Phi(R_{ij})$ is a rearrangement potential at a constant volume, with the volume largely fixed by the f(V) term in Eq.(3), so that the equilibrium nearest neighbour distance between atomic center does not necessarily lie at or near any minimum of $\Phi(R_{ij})$.

The theory of Eq.(3) is valid even for atoms containing *d*-electrons as shown by Caroli [37] provided one adds an appropriate resonance term [29] to $\Phi(R_{ij})$. In this form the theory has been applied successfully to the alkaline earth metals by various authors ( see for example References [38–41] ).

We now apply Eq.(3) to be the variation of the total energy with *c/a* ratio. We consider the system at fixed volume $V_{at}$ per atom

$$V_{at} = \frac{\sqrt{3}}{2}a^2c = Ca^3\gamma = Ca_0^3\gamma_0, \qquad (4)$$

where $\gamma$ is the *c/a* ratio and $\gamma_0 = \sqrt{8/3} \approx 1.632$, the ideal *c/a* ratio. Also C is a constant, *a* is the usual lattice constant and $a_0$ is the lattice constant for the ideal structure with the same atomic volume. The non-ideal structure has six nearest neighbours at a distance *a* in the close packed plane perpendicular to the *c* axis, and six nearest neighbours at a distance *d*, say, in the close packed layers just above and below [42]. The total sum of pairwise interactions over the nearest neighbours can therefore be written ( per atom ) as

$$\Phi_{tot} = \frac{1}{2}[6\Phi(a) + 6\Phi(d)]$$
$$= 6\Phi(a_0)$$

$$+3 \times (\frac{\partial \Phi}{\partial r}|_{r=a_0} (a-a_0) + \frac{1}{2}\frac{\partial^2 \Phi}{\partial r^2}|_{r=a_0} (a-a_0)^2 + \frac{1}{3!}\frac{\partial^3 \Phi}{\partial r^3}|_{r=a_0} (a-a_0)^3 + ...)$$
$$+3 \times (\frac{\partial \Phi}{\partial r}|_{r=a_0} (d-a_0) + \frac{1}{2}\frac{\partial^2 \Phi}{\partial r^2}|_{r=a_0} (d-a_0)^2 + \frac{1}{3!}\frac{\partial^3 \Phi}{\partial r^3}|_{r=a_0} (d-a_0)^3 + ...). \tag{5}$$

It is convenient to define the derivation $\eta$ from ideality by

$$\eta = 1 - \frac{\gamma}{\gamma_0}, \tag{6}$$

and then the $a$, $d$ can be expanded in terms of the $\eta$ as follows:

$$a = a_0(1 - \frac{1}{3}\eta + \frac{2}{9}\eta^2 - \frac{14}{81}\eta^3 + ...),$$
$$d_{nn} = a_0(1 + \frac{1}{3}\eta + \frac{1}{9}\eta^2 - \frac{11}{81}\eta^3 + ...). \tag{7}$$

Substituting into Eq.(5) gives

$$\Delta\Phi_{tot}(\eta) = \Phi_{tot} - \Phi_{tot}(ideal)$$
$$= (\frac{\partial \Phi}{\partial r}|_{r=a_0} + \frac{1}{3}a_0\frac{\partial^2 \Phi}{\partial r^2}|_{r=a_0})a_0\eta^2 + (\frac{25}{27}\frac{\partial \Phi}{\partial r}|_{r=a_0} + \frac{1}{9}a_0\frac{\partial^2 \Phi}{\partial r^2}|_{r=a_0})a_0\eta^3 + ... \tag{8}$$

As might be expected, the linear term in $\eta$ cancels out, and the stability or otherwise of the ideal structure is largely determined by the sign of the second order term. If it goes negative, then we expected the equilibrium $c/a$ to derivate substantially from the ideal value. The third order term determines whether the equilibrium will occur at $c/a$ greater or less than ideal value ( $\sqrt{8/3}$ ).

We have taken $\Phi(R_{ij})$ for barium from the work of Jank and Hafner [29] which included both the $sp$ and $d$-electron contributions at one volume corresponding to ambient pressure. We have not recalculated $\Phi(R_{ij}, V)$ at smaller and larger proportion of $d$-electrons or $\Phi(R_{ij}, V)$ at smaller volumes corresponding to high pressures for several reasons. Firstly $\Phi(R_{ij}, V)$ in general does not depend strongly on volume [22] [36] although in the present case the variation in the proportion of $d$-electrons will effect it. Secondly we are seeking here a semiquantitative understanding of the experimental results, not trying to reproduce the data accurately. Thirdly at the pressures considered here, there may be an appreciable additional interaction due to the $5p^6$ core shell. Fourthly in barium under pressure the proportion of $d$-electrons has become so large that we are outside the proper range of validity of the theory: in fact there will be significant 3-atom *etc.* contributions from the $d$-band. However we believe the theory with a fixed $\Phi(R_{ij})$ captures the dominant terms in the energy and suffices to give a correct interpretation of the experimental results.

The $\Phi(R)$ taken from Ref. [29] is shown in Fig.5, and also shown broken down into separate $s$ and $d$ contribution. The nearly-free-electron tight-binding-bond ( NEF-TBB ) approach around a band width of $W_d = 7.0eV$ and an average nearest neighbour $d$-$d$ hopping integral of $h(R_0) = 0.6217eV$. A configuration of 1.25 $s$- electron and 0.75 $d$-electron was assumed. The $\Phi(R)$ has the customary form, a very repulsive screened pseudo-ion potential at short range, going over to Friedel oscillation at larger distance. The total $\Phi(R)$ from $s$- and $d$-electron could be fitted quite well, especially in the region of 3 to $4\mathring{A}$ of interest here, by the form

$$\Phi(R) = C\frac{\cos(C_1 R)}{(C_1 R)^\delta} + \frac{B}{(R/R_0)^{12}} \tag{9}$$

with the parameter C=-35.75 Ry, $C_1 = 1.15 \mathring{A}^{-1}, \delta = 4.32$, B=1.90 mRy and $R_0 = 3.1 \mathring{A}$. An analytic fit was necessary for the subsequent calculations.

It is clear from Eq.(8) that the instability of the ideal structure is driven by a cancellation between a negative first derivative $\Phi'$ and a positive second derivative $\Phi''$. The negative part of the Friedel oscillation tends to cancel the positive tail of the hard core repulsion, resulting in a very steep $\Phi(R)$ and large negative $\Phi'$ [36]. In our case the attractive part of the $d$-electron interaction has the same effect as seen in Fig.5, making a large contribution in the range $R = 3.0$ to $3.5\mathring{A}$. It is this steepness in $\Phi(R)$ which drives the instability, i.e. large negative $\Phi'$ in Eq.(8).

Using the fitted $\Phi(R)$, we have evaluated $\Phi_{total}$ from the 12 nearest neighbours as a function of $c/a$ at various fixed volumes ( corresponding in Fig.6 and 7 to fixed $a_0$ as defined by Eq.(4) ). Figure 6 shows that an instability of the ideal structure does indeed set in with decreasing volume. This is taken to represent the experimentally observed situation in phase $II$. The calculated onset at $a_0 = 3.25 \mathring{A}$ does not correspond exactly the the observed onset at $a_0 \approx 4 \mathring{A}$ ( Fig. 1 inset ), but such a difference only to the expected in view of the various approximation made. In fact that curves of the $s$ and $d$ component in Fig.5 cross at $R = 2.6 \mathring{A}$ with the $d$ component being very much the harder one. Thus the greater number of $d$-electrons in barium under pressure ( 1.2 to 1.5, Fig.2 ) compared with the number 0.75 assumed in Ref. [29], will move the onset of instability to larger values of $a_0$ as required by the data, as shown by the left arrow in the inset of Fig.5. The calculations were carried out with the expansion of Eq.(8) and not with $\Phi(R)$ directly so that no energy minimum is obtained at $a_0 < 3.25 \mathring{A}$. However the asymmetry of the curves shows clearly that the absolute minimum will lie at $c/a$ less than the ideal value, as observed in barium.

Further similar total energy calculations at constant volume in the lower range of $a_0 = 2.1$ to $2.6 \mathring{A}$ are shown in Fig.7. On decreasing the volume, the instability now disappears so that the equilibrium structure again become close to the ideal packing. This corresponds to the observation in phase $V$. In the calculation the reversion to ideal close packing occurs at $a_0 = 2.3 \mathring{A}$, again too low a value compared with experiment for the same reason as before. Since there is no linear term in $\eta$ in Eq.(8), the $c/a$ will be exactly ideal once the system is out of the instability in our simple theory. In reality there is a small linear term from more distant neighbours and other higher order terms in the theory, which are beyond the present calculations. A reversion to " hard sphere " behaviour and hence ideal packing is expected at very high pressure because atoms become harder and harder at small distance. It is easy to verify from Eq.(8) that a simple interaction

$$\Phi(R) \sim \exp(-\frac{R}{\lambda}) \tag{10}$$

gives the ideal structure as stable for $\lambda$ less than $a_0/3$, i.e. less than $1\mathring{A}$ in our case. Even the Thomas Fermi screening length from the conduction electrons is of the order of $0.5\mathring{A}$, and the effect of the $d$-electrons representing the cut-off of the atomic $d$-orbital function will be much sharper, as shown by the right arrow in the inset of Fig.5.

## V. CONCLUSIONS

In conclusion, we have assembled various types of calculation to give an overall understanding of the observed behaviour of barium in the two $hcp$ structures, phase $II$ and $V$, under pressure. Firstly there is the electron transfer, mainly $s$ to $d$ as expected from previous works, but saturating and with some transfer into $f$ orbitals at highest pressures ( 50–100GPa ) in phase $V$. Fully self-consistent FP-LMTO total energy calculations have reproduced the observed behaviour of the $c/a$ ratio quite well, and by dropping various orbitals from the basis set we have demonstrated that it is indeed the transfer into $d$-orbitals that is driving the instability in $c/a$ in phase $II$, and the fact that $c/a$ is less than the ideal value in both phases $II$ and $V$. Finally we have discussed the instability of the ideal structure in phase $II$ and its re-establishment in phase $V$ in terms of a simple pseudopotential perturbation theory, as modified by Hafner and other authors to take into account the influence of the $d$-component in the electron states. This theory was taken over from a similar discussion relating to the behaviour of the $c/a$ ratio in the series Be, Mg, Zn, Cd, and Hg [43,44] where there is an instability of the ideal close-packed structure similar to that in phase $II$ of barium. Several approximation mean that the theory is only a semiquantitative interpretation of the experimental data. However we expect that picture it presents to be valid, because experience with many such pseudopotential perturbation calculation has shown that they tend to be quite robust, i.e. not sensitive to detailed tuning of the pseudopotential [45], with the second order terms really being dominant [46]. The theory involves the pairwise interatomic rearrangement potential $\Phi(R_{ij}, V)$ at constant volume. The instability of the ideal structure is driven by $\Phi$ being very steep where a very repulsion screened ion interaction at short R goes over into an attractive outer region due to the $d$-electrons and Friedel oscillations.

## ACKNOWLEDGMENTS

Useful discussions with A.I.Liechtenstein about the FP-LMTO program are gratefully acknowledged. W.-S.Zeng would like to acknowledge financial support by the Max-Planck-Gesellschaft during his stay at the Max-Planck-Institut für Festkörperforschung, Stuttgart, Germany. W.-S.Zeng gratefully acknowledge O.K.Andersen for his great helps and many encouragements.

# FIGURE CAPTIONS

Fig.1. The experimental structure data of barium under high pressure. The lattice constants $a$ and $c$ are shown in the inset.

Fig.2. The LMTO-ASA calculated occupation numbers of $s$, $p$, $d$ and $f$ orbitals as functions of pressure.

Fig.3. Total energy (relative) of barium in the $hcp$ structure in phases $II$ and $V$ as a function of the $c/a$ ratio, calculated with the FP-LMTO method with a large basis set. Each curve is displaced vertically by an arbitrary amount for clarity. The arrows indicate the $c/a$ ratios found experimentally.

Fig.4. Total energy (relative) of barium in the $hcp$ structure (a) in phase $II$ at a volume corresponding to 7.21 GPa and (b) in phase $V$ for 57.73 GPa as a function of the $c/a$ ratio for different choices of the basis set. Each curve is displaced vertically by an arbitrary amount for clarity, labelled according to the notation explained in the text.

Fig.5. The pair-potential of barium as a function of interatomic distance from Ref. [29] using hybridized pseudopotential tight-binding-bond theory, showing the separate contribution from the $s$- and the $d$-electrons, the total $s+d$ and the fit of Eq.(9) to the $s+d$ curve.

Fig.6. The pair potential contribution from the 12 nearest neighbours to the total energy of barium as a function of $c/a$ at various fixed volumes ( indicated by the respective $a_0$ ), relative to the ideal structure. The behaviour is similar to that observed in phase $II$, the ideal structure becoming unstable at lower volume.

Fig.7. The same as Fig.6 but for smaller volumes. The behaviour is that observed in phase $V$, the ideal structure becoming stable again.

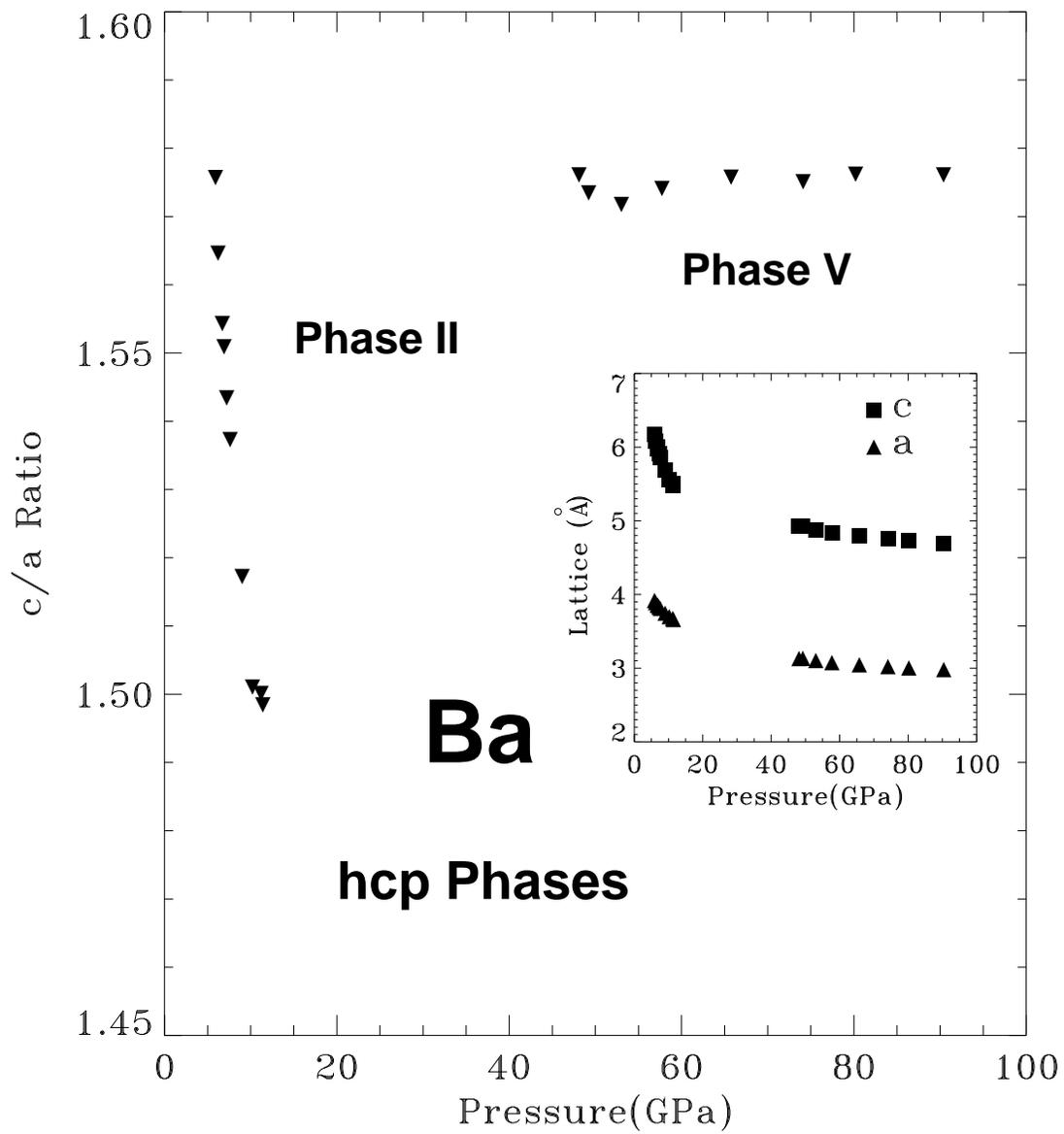

Figure 1

W.-S.Zeng et al

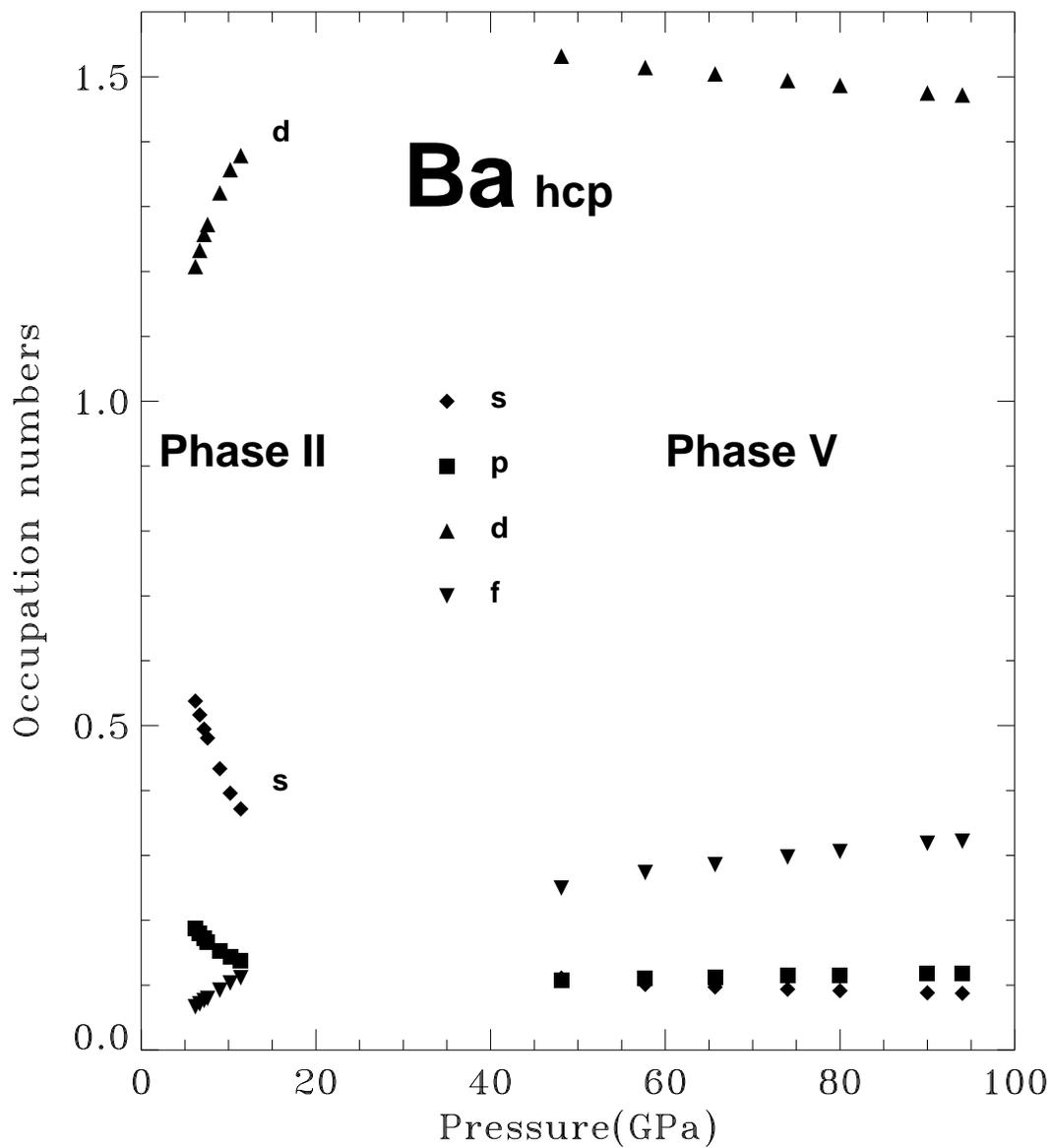

Figure 2

W.-S.Zeng et al

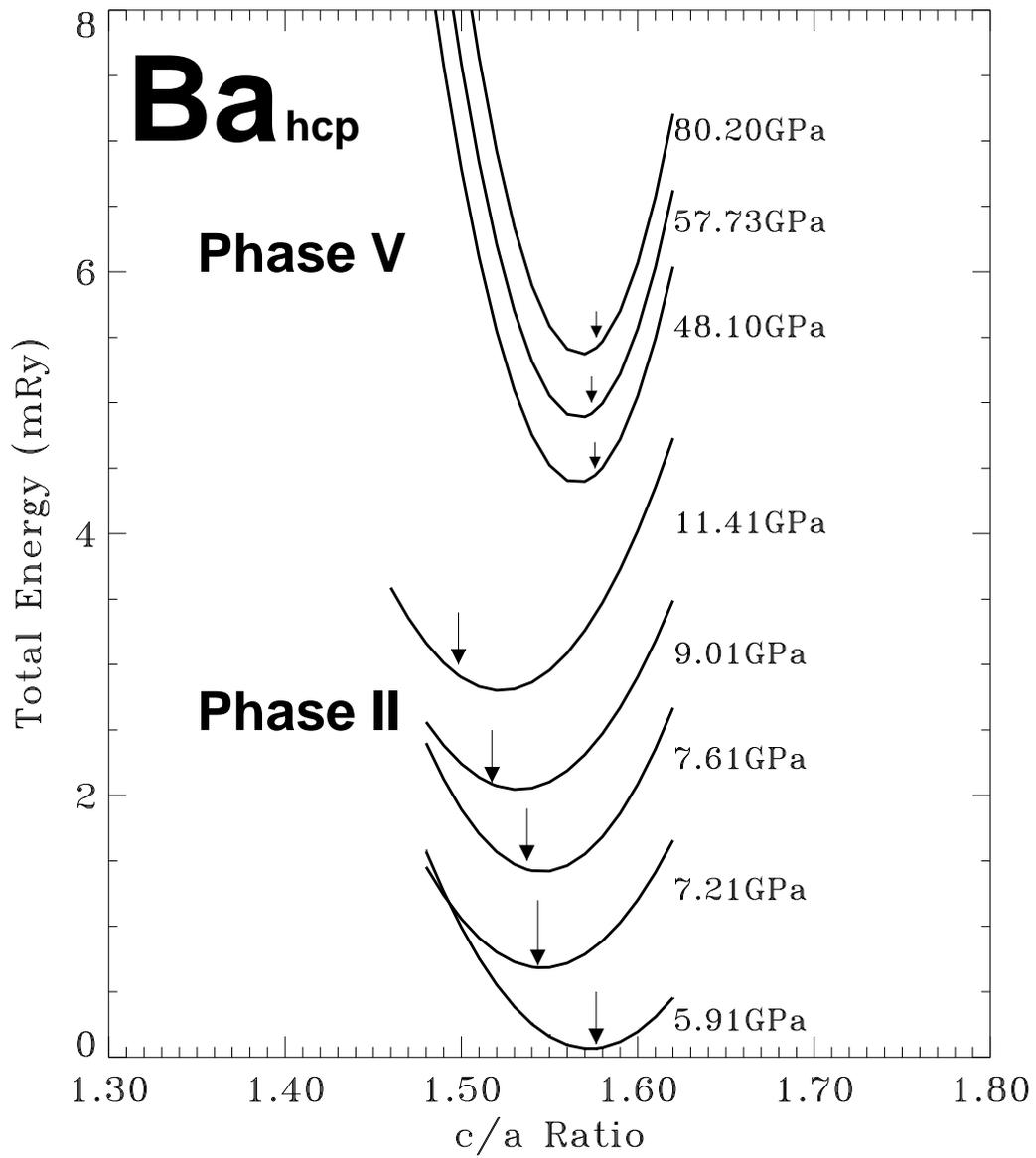

Figure 3

W.-S.Zeng et al

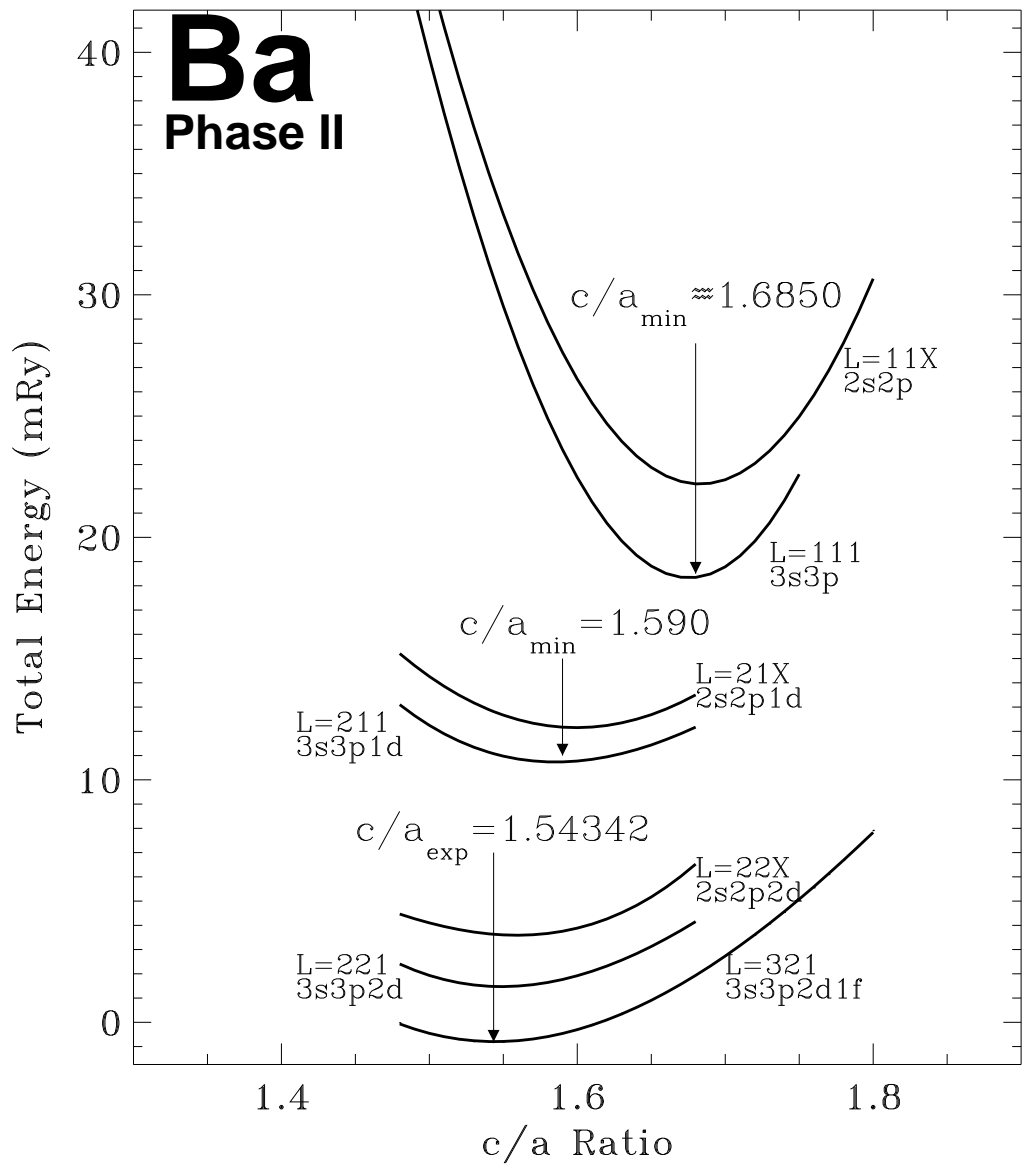

Figure 4a

W.-S.Zeng et al

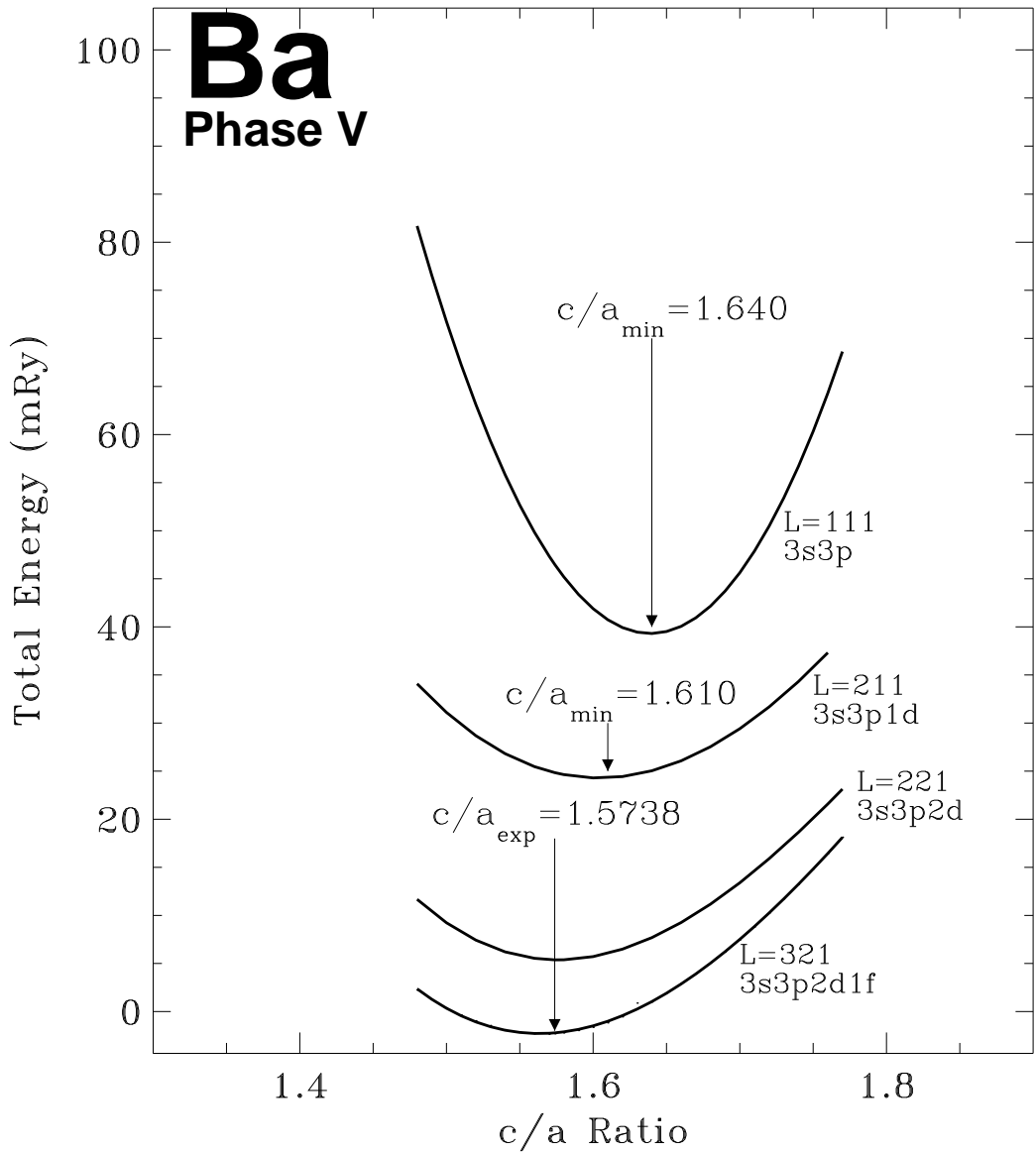

Figure 4b

W.-S.Zeng et al

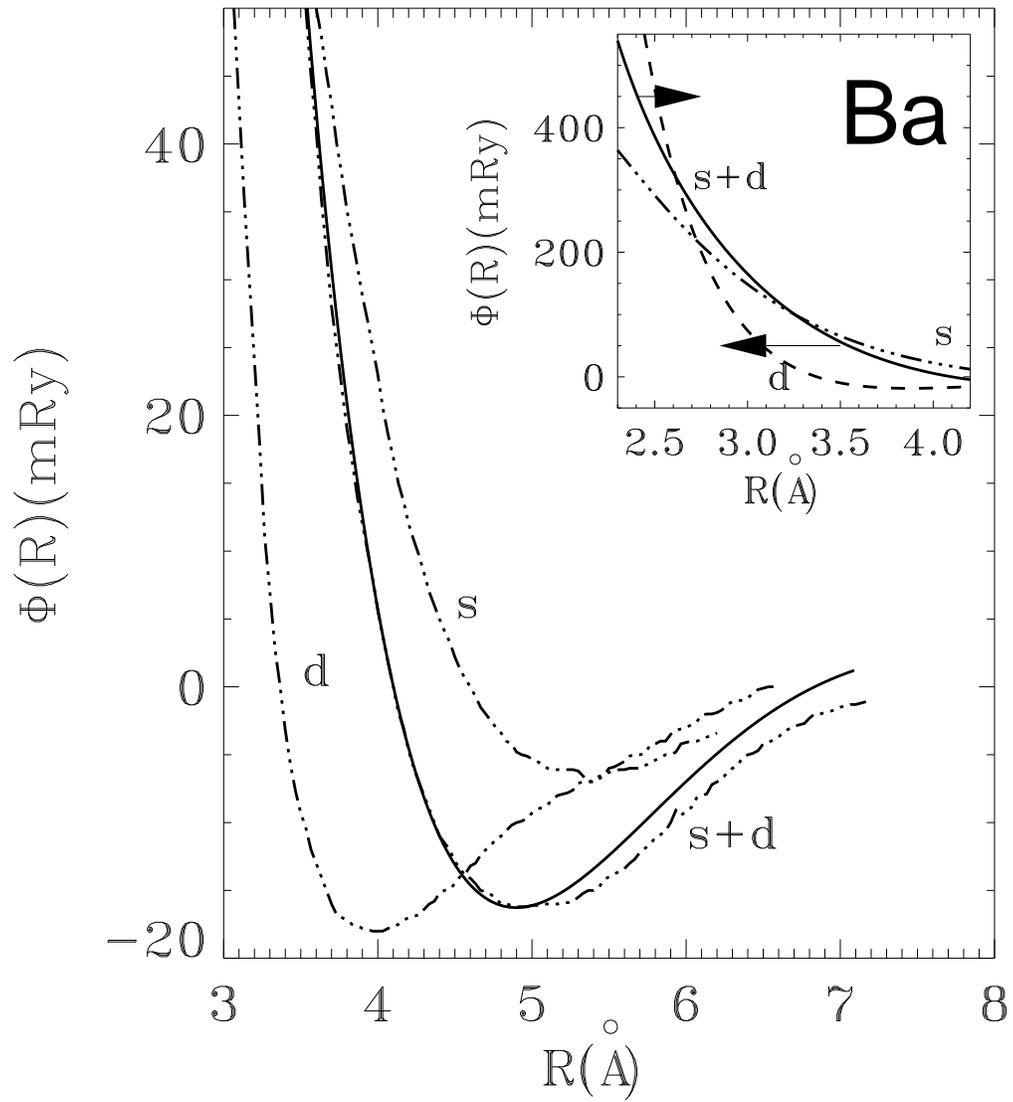

Figure 5

W.-S.Zeng et al

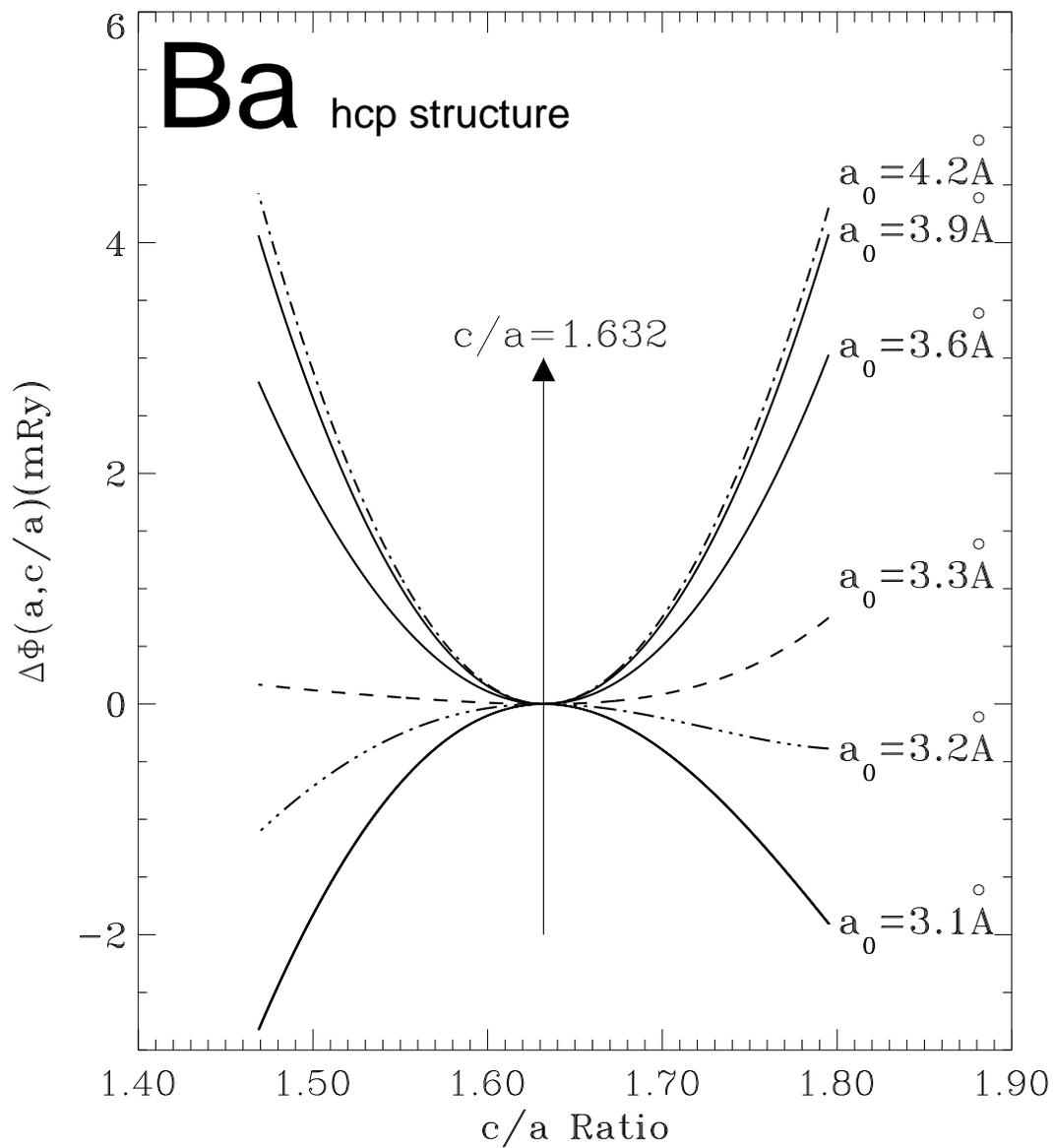

Figure 6

W.-S.Zeng et al

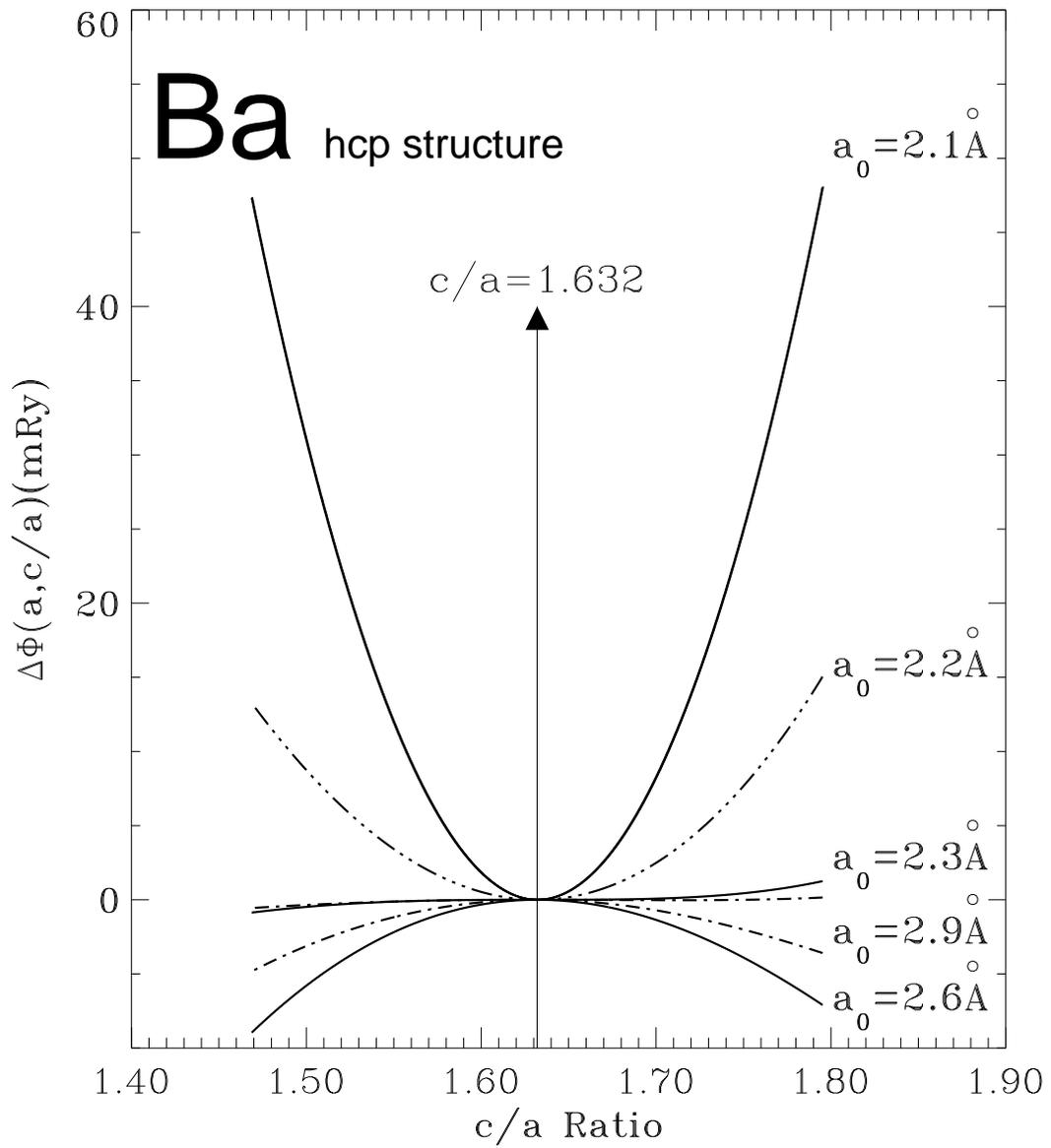

Figure 7

W.-S.Zeng et al